\documentclass[aps,pra,10pt,superscriptaddress,twocolumn,nofootinbib]{revtex4-2}
\usepackage{graphicx}
\usepackage{xcolor,soul}
\usepackage{orcidlink}
\usepackage{amsmath,amsthm,amssymb}
\usepackage{mdframed}
\usepackage{hyperref}
\newcommand{\ket}[1]{\left| #1 \right\rangle}
\newcommand{\bra}[1]{\left\langle #1 \right|}
\begin{document}

\title{Counterfactual restrictions and Bell's theorem}
\author{Jonte R. Hance\,\orcidlink{0000-0001-8587-7618}}
\email{jonte.hance@newcastle.ac.uk}
\affiliation{School of Computing, Newcastle University, 1 Science Square, Newcastle upon Tyne, NE4 5TG, UK}
\affiliation{Quantum Engineering Technology Laboratories, Department of Electrical and Electronic Engineering, University of Bristol, Woodland Road, Bristol, BS8 1US, UK}

\begin{abstract}
    We show that the ability to consider counterfactual situations is a necessary assumption of Bell's theorem, and that, to allow Bell inequality violations while maintaining all other assumptions, we just require certain measurement choices be counterfactually restricted, rather than the full removal of counterfactual definiteness. We illustrate how the counterfactual definiteness assumption formally arises from the statistical independence assumption. Counterfactual restriction therefore provides a way to interpret statistical independence violation different to what is typically assumed (i.e. that statistical independence violation means either retrocausality or superdeterminism).  We tie counterfactual restriction to contextuality, and show the similarities to that approach.
\end{abstract}

\maketitle
\section{Introduction}

Counterfactual definiteness is the intuitive idea that it is meaningful to consider what the values of any currently-defined observable, or the result of a given measurement, would instead be, if the world was different in some specific way. This forms part of our day-to-day experience; we expect a small change to affect the universe in a measurable way. Quantum mechanics casts doubt on this. Standard quantum mechanics has been described as acting as though unperformed experiments have no results \cite{Peres2004QIandRel}, making it meaningless to discuss the impact of a small change in circumstances on a result. In other words, quantum mechanics has been said to require counterfactual indefiniteness.

By looking in more detail at each of the assumptions required to make a Bell inequality, we show that counterfactual definiteness (or the matching condition) is a necessary assumption to generate Bell inequalities, but which appears formally in the derivation of these inequalities as part of the statistical independence (or measurement independence) condition. We show therefore that the statistical approach to generating Bell inequalities does not avoid this mathematical assumption, despite what is often claimed. We further show that, contrary to how we normally think of a violation of statistical independence as only coming from superdeterminism or retrocausality, a violation of the counterfactual definiteness aspect of the statistical independence condition can also come from a restriction on those counterfactual scenarios which are nomologically possible (something we term ``a counterfactual restriction'').

This paper is laid out as follows. In Section \ref{sect:EPRBell}, we discuss Bell's Theorem, starting in Section \ref{subsect:EPR} by going over the EPR paradox, laying out how to generate (and violate) a Bell inequality (specifically the CHSH inequality) in this scenario in Section \ref{subsect:bell}. Section \ref{sect:Assumpts} looks at the assumptions required to generate a Bell inequality. Section \ref{subsect:HV} describes the assumption of hidden variables, and Section \ref{subsect:Factorisability} examines the assumption of factorisability, which is formed of the intersection of No Superluminal Interaction (Section \ref{subsect:NoSuperluminal}) and Statistical Independence (Section \ref{subsect:SI}). In Section \ref{subsect:Comparison} {we} then look {at} an assumption whose inclusion is often debated: that we can compare measurement results from different trials. We look in more detail at this comparison assumption in Section \ref{sect:CFRestrict}, showing in Section \ref{subsect:CFRnotCFI} that Bell inequality violation only requires counterfactual restriction, rather than a full removal of counterfactual definiteness; in Section \ref{subsect:CFRasSIV} that this counterfactual restriction links to, and provides another interpretation of, the violation of Statistical Independence; and in Section \ref{subsect:Contextuality} that this interpretation links naturally to contextuality. We summarise our argument in Section \ref{sect:Conc}.

\section{EPR and Bell's Theorem}\label{sect:EPRBell}

 In this section, we go over Bell's Theorem. This states that, given certain (reasonable-sounding) classical assumptions, one can generate an inequality from the expected results of measurements in an experimental scenario, which can be violated by expected results of measurements in this scenario when represented using quantum mechanics \cite{bell1971introduction}. The idea of specifically considering cases where quantum mechanics gives differing predictions to those expected from reasonable assumptions about the nature of the world, originated with Einstein, along with Podolsky and Rosen (EPR) \cite{EPR}. They first came up with a scenario in which interpreting quantum mechanics as fully describing a given situation seemingly leads to a paradox. In this section, we first go through this EPR paradox, and Bohm's simplification of it, before showing how to use this EPR-Bohm set-up to derive a Bell inequality: the CHSH inequality. We then describe experimental attempts to violate this inequality (e.g. Freedman and Clauser's \cite{Freedman1972}, and Aspect et al's \cite{Aspect1981BellViol,Aspect1982}), loopholes in those experiments, and how those loopholes were resolved. This serves as the groundwork necessary for the discussion in the next Section of the assumptions necessary to form a Bell inequality.

\subsection{The EPR Paradox}\label{subsect:EPR}

Einstein, Podolsky and Rosen came to a paradox by combining two key concepts in quantum mechanics: conjugacy (one variable's uncertainty increasing as the other's decreases) and entanglement (two particles' states being so correlated they cannot be written independently) \cite{EPR}.

When we measure one of a set of conjugate variables for one of a pair of entangled particles, quantum mechanics says the conjugate variable for the other particle becomes uncertain. This is despite the two particles potentially being spatially separated, and standard quantum mechanics providing no physical mechanism for a signal to propagate between them.

\begin{figure}
    \centering
    \includegraphics[width=\linewidth]{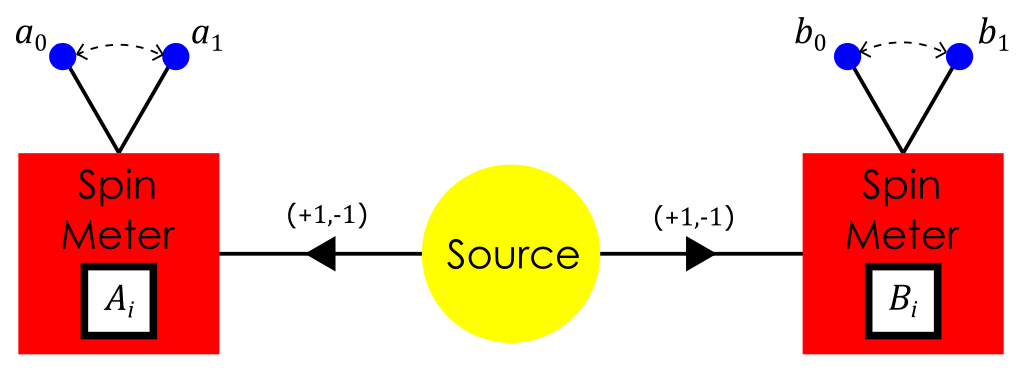}
    \caption{The EPR-Bohm experiment \cite{bohm1951quantum}, where a source emits pairs of spin-entangled particles in opposite directions. The joint spin state of the two particles is $\ket{\Psi}=(\ket{\uparrow}_A\otimes\ket{\downarrow}_B + \ket{\downarrow}_A\otimes\ket{\uparrow}_B)/\sqrt{2}$. The left-hand particle (particle $A$) is subjected to either test $a_0$ or $a_1$, and the right-hand particle (particle $B$), is subjected to either test $b_0$ or $b_1$. All tests have result either $+1$ or $-1$. This set-up (alongside various assumptions, discussed below) was used to derive the Clauser-Horne-Shimony-Holt (CHSH) inequality \cite{CHSH}, which was experimentally violated by Aspect et al \cite{Aspect1982}.}
    \label{fig:BohmBellTest}
\end{figure}

Bohm simplified this scenario, doing away with non-commuting observables, and focusing on joint states---those which quantum mechanics says cannot be written as the (tensor) product of single-particle states \cite{bohm1951quantum}. He gave the example of the joint spin state of two entangled electrons, where the total spin of the two particles is 0 (see Fig \ref{fig:BohmBellTest}).

Using bra-ket notation \cite{dirac1939new}, we write the spin-states of the two electrons as the {(spin-singlet \mbox{\cite{Stuckey2020Mermin}})} Bell state
\begin{equation}
\label{Eq:Bohm}
    \ket{\Psi^-}=\frac{1}{\sqrt{2}}(\ket{\uparrow}_A\otimes\ket{\downarrow}_B - \ket{\downarrow}_A\otimes\ket{\uparrow}_B)
\end{equation}
where indices $A$ and $B$ label the spin states of particles $A$ and $B$ respectively. Note, while normally we need to define a basis for the result of a spin measurement (e.g., for a spin-1/2 particle, spin-up or spin-down in the $X$, $Y$ or $Z$ direction), the Bell state in Eq.~\ref{Eq:Bohm} takes the same form regardless of the basis in which we represent it, so long as we represent both particles in the state in the same basis---i.e.,
\begin{equation}
    \begin{split}
        \ket{\Psi^-} &= \frac{1}{\sqrt{2}} (\ket{\uparrow_X}_A\otimes\ket{\downarrow_X}_B - \ket{\downarrow_X}_A\otimes\ket{\uparrow_X}_B)\\
        &= \frac{1}{\sqrt{2}} (\ket{\uparrow_Y}_A\otimes\ket{\downarrow_Y}_B - \ket{\downarrow_Y}_A\otimes\ket{\uparrow_Y}_B)\\
        &= \frac{1}{\sqrt{2}} (\ket{\uparrow_Z}_A\otimes\ket{\downarrow_Z}_B - \ket{\downarrow_Z}_A\otimes\ket{\uparrow_Z}_B)
    \end{split}
\end{equation}
(and similarly for any other choice of representation direction).

As this state cannot be written as the tensor product of the two individual electrons' states, quantum mechanics says we cannot describe these two electrons' spins independently of one another. Given the collapse postulate, if the description in Eq.~\ref{Eq:Bohm} completely describes the physical reality of the scenario, then when one of the electrons' spins is measured, it somehow collapses the overall state, pushing the other electron into its corresponding spin (and making its spin in any of the other two mutually unbiased bases uncertain). If we are being fully realist about both the wavefunction (i.e. saying there is nothing in this system not described by the wavefunction as given/no hidden variables) and this collapse, this happens instantaneously, and so information must travel superluminally. To reinforce this, the state into which the first electron collapses is completely random according to standard quantum mechanics, so there is no way (by this account) that the second electron can be said to be `pre-prepared' into the corresponding state in advance, to avoid the need for superluminal information transfer. 

Einstein dismissed this effect as ``spooky action(s) at a distance" (spukhafte Fernwirkung{en}) \cite{Einstein1947Letter}---information seemingly passing from one place to another, instantly and without mechanism. To him, this meant there must be a deeper description of what was happening in the system than standard quantum mechanics provided---that there was some ``hidden variable'', not described by the formalism of quantum mechanics (and so missing from Eq.~\ref{Eq:Bohm}) which governed the underlying situation in order to avoid this superluminal update. Models which describe this possibility are referred to as ``hidden-variables'' models \cite{adlam2023taxonomy}.

\subsection{Bell's Theorem}\label{subsect:bell}

For nearly thirty years after the EPR paper, whether you believed quantum mechanics was incomplete, failing to account for some `hidden variables' which further explain its peculiarities, or instead violates some fundamental assumption about the nature of the universe, was just considered a matter of interpretation. However, in 1964 Bell proposed an experiment to identify testable differences between these two options \cite{Bell1964}. For the EPR-Bohm experiment, Bell derived the upper limit for measurable correlations between the two particles, assuming they obeyed a local hidden-variable model (among other assumptions).

These local hidden variable models posit a variable $\lambda$, which, from the moment an entangled state is generated, holds information as to the final state its constituents will end up in, above and beyond that given by the wavefunction. For instance, for Eq.~\ref{Eq:Bohm}, this information would be whether the spins of the two electrons will collapse to $\ket{\uparrow}_1\otimes\ket{\downarrow}_2$ or $\ket{\downarrow}_1\otimes\ket{\uparrow}_2$). This variable is hidden (i.e., not necessarily accessible to the experimenter), and local (i.e., cannot be used as a channel to superluminally transmit information between the two particles). The space $\Lambda$ spans all possible values for this local hidden variable (i.e. $\lambda\in\Lambda,\;\forall\lambda)$. (We go into more detail about these assumptions in Section \ref{subsect:HV}.)

By using properties of these local hidden variable models alongside certain assumptions, Bell (and others) derived inequalities for sums of correlation functions between measurement results. The most famous of these Bell inequalities is the Clauser-Horne-Shimony-Holt (CHSH) inequality \cite{CHSH} (of which we use the Wigner form \cite{Wigner1970Bell,redhead1987incompleteness}). As in the EPR-Bohm set-up, a source emits pairs of particles in opposite directions. The left-hand particle is subjected to test $a$ (either $a_0$ or $a_1$), and the right-hand one to test $b$ (either $b_0$ or $b_1$). All tests have result either $+1$ or $-1$ (see Fig.~\ref{fig:BohmBellTest}). For example, for tests of single-photon polarisation, where $a_0$ and $a_1$ ($b_0$ and $b_1$) are differently oriented polarisers, +1 corresponds to the photon being transmitted, and -1 to it being absorbed.

For a given trial $i$, we call the result $A_i$ of the experiment of the left either $A_{0,i}$ or $A_{1,i}$ (depending on whether we choose test $a$ to be $a_0$ or $a_1$ for that trial), and similarly $B_i$ as either $B_{0,i}$ or $B_{1,i}$. Therefore, we can write the four pairs of products of the results as $A_{0,i} B_{0,i}$, $A_{1,i} B_{0,i}$, $A_{0,i} B_{1,i}$, and $A_{1,i} B_{1,i}$.

We use these 4 result combinations to give
\begin{equation}
    \label{CHSH22}
    \begin{split}
    &\gamma_i(a_0,a_1,b_0,b_1)=\\
    &A_{0,i} B_{0,i}+A_{1,i} B_{0,i}+A_{0,i} B_{1,i}-A_{1,i} B_{1,i}
    \end{split}
\end{equation}

As $A_{0,i},A_{1,i},B_{0,i},B_{1,i}\in\{-1,1\}$,
 \begin{equation}
 \begin{split}
      \gamma_i(a_0,a_1,b_0,b_1)&=A_{0,i}(B_{0,i}+B_{1,i})+A_{1,i}(B_{0,i}-B_{1,i})\\
      &=\pm 2
 \end{split}
 \end{equation}

 For $N$ events, given $\gamma_i=\pm 2$,
 \begin{equation}\label{eq:CHSHFiniteN}
 \begin{split}
     \left| \frac{1}{N}\sum^N_{i=1}\gamma_i\right| = \Bigg| \frac{1}{N}\sum^N_{i=1}A_{0,i} B_{0,i} +  \frac{1}{N}\sum^N_{i=1}A_{1,i} B_{0,i}\\
     +  \frac{1}{N}\sum^N_{i=1}A_{0,i} B_{1,i} -  \frac{1}{N}\sum^N_{i=1}A_{1,i} B_{1,i} \Bigg| \leq 2
\end{split}
 \end{equation}

 As $A_i \in \{A_{0,i}, A_{1,i}\}$ and $B_i \in \{B_{0,i}, B_{1,i}\}$, the expectation value for the product of the results of a given pair of measurements $a,b$ is
 \begin{equation}
     E(a,b) = \lim_{N\rightarrow\infty} \frac{1}{N}\sum^N_{i=1}A_iB_i
 \end{equation}
so Eq.~\ref{eq:CHSHFiniteN} becomes
\begin{equation}
\begin{split}
    S &= \left| E(a_0,b_0) + E(a_1,b_0) + E(a_0,b_1) - E(a_1,b_1)\right|\\
    &\leq 2
\end{split}
\end{equation}

This is the CHSH inequality. If, for any set of measurement settings a quantum version of the system can get a value of $S$ greater than $2$, then this shows that quantum mechanics must violate one of the assumptions we used to create our Bell inequality, and gives us a way of experimentally testing whether or not the universe follows all of these assumptions. Note, the CHSH inequality is typically used as, unlike Bell's 1964 inequality, it doesn't require perfect anticorrelation between the measurement outcomes of the two particles when measured using the same setting. This allows it to better account for experimental errors; with perfect anticorrelation, it simplifies to Bell's initial inequality \cite{bell1971introduction}.

Quantum mechanics can violate this limit. Returning to the state defined in Eq.~\ref{Eq:Bohm}, the quantum mechanical expectation value for the product of the results of two tests $a,b$ on the two parts of the entangled state (i.e., test $a$ on the part of the state labelled subscript 1, and test $b$ on the part labelled subscript 2) is
\begin{equation}\label{Eq:ExpectQuantum}
    E(a,b) = \bra{\Psi}(\sigma_{1,a}\otimes\sigma_{2,b})\ket{\Psi}
\end{equation}
where $\sigma_{1,a}$ is the spin operator on particle 1 in direction $a$, and $\sigma_{2,b}$ the spin operator on particle 2 in direction $b$, such that for instance
\begin{equation}
 \sigma_{1,Z}\ket{\uparrow_Z}_1 = \ket{\uparrow_Z}_1;\; \sigma_{1,Z}\ket{\downarrow_Z}_1 = -\ket{\downarrow_Z}_1
\end{equation}

As spin lives on the Bloch sphere, we can simplify Eq.~\ref{Eq:ExpectQuantum} by decomposing spin measurements in the $b$ direction, into the component of the measurement parallel to $a$, and the component perpendicular to $a$ (denoted $a\perp$):
\begin{equation}\label{Eq:ExpectQAngle}
\begin{split}
    &E(a,b) = \\
&\bra{\Psi}\left(\sigma_{1,a}\otimes\left({\cos\theta_{ab}}\sigma_{2,a} + \sin{\theta_{ab}}\sigma_{2,a\perp}\right)\right)\ket{\Psi}
    \end{split}
\end{equation}
where $\theta_{ab}$ is the angular distance along the Bloch sphere between $a$ and $b$.

As $\ket{\Psi}$ is the state given in Eq.~\ref{Eq:Bohm}, which keeps the same form regardless of the basis we represent it in, Eq.~\ref{Eq:ExpectQAngle} simplifies further to
\begin{equation}
\begin{split}
 E(a,b)= -\cos{\theta_{ab}}
\end{split}
\end{equation}

This means, given
\begin{equation}
        S = \left| E(a_0,b_0) + E(a_1,b_0) + E(a_0,b_1) - E(a_1,b_1)\right|
\end{equation}
in the quantum case,
\begin{equation}
    S = \left|-\cos{\theta_{a_0b_0}}-\cos{\theta_{a_1b_0}}-\cos{\theta_{a_0b_1}}+\cos{\theta_{a_1b_1}}\right|
\end{equation}

This is maximal when $\theta_{a_0b_0} = \theta_{a_1b_0} = \theta_{a_0b_1} = \pi/4$ and $\theta_{a_1b_1} = 3\pi/4$, such as when 
\begin{equation}
    a_0 = \sigma_X;\;a_1=\sigma_Z;\;b_0 = \frac{\sigma_Z+\sigma_X}{\sqrt{2}}\;\;b_1 = \frac{\sigma_Z-\sigma_X}{\sqrt{2}}
\end{equation}
%Could add graph here, showing all four measurement operators on Bloch sphere?
and gives
\begin{equation}
    S_{max} = \left| -\frac{1}{\sqrt{2}}-\frac{1}{\sqrt{2}}-\frac{1}{\sqrt{2}}-+\frac{1}{\sqrt{2}}\right| = 2\sqrt{2}
\end{equation}

This is a proof of the Tsirelson bound \cite{cirel1980quantum}: the maximal value of $S$ achievable using quantum mechanics. (Note, for the EPR-Bohm scenario, Popescu and Rohrlich showed $S$ can go above this bound, up to $S\leq4$, for general nonlocal theories \cite{Popescu1994Boxes}, though Carmi and Cohen showed models where $S>2\sqrt{2}$ must at the least violate a subtle form of relativistic causality \cite{Carmi2019Popescu}). While there are more formal proofs, generalised for all quantum systems (rather than just dimension-2, like the spin-1/2 qubits above), this proof serves our purpose. It shows the CHSH inequality can be violated using the quantum formalism: local-hidden-variable models of the sort used to generate the CHSH inequality above fail to account for predicted quantum correlations between the two particles. Therefore, an experiment giving $S>2$ would show the Universe violates one of the assumptions used to form this inequality. While not necessary for our argument, in Appendix~\ref{sec:Exp}, we discuss the history of experiments used to demonstrate values of $S>2$, loopholes in those experiments, and how those loopholes were eventually closed.

\section{Formal Assumptions and the Consequences of their Violation}\label{sect:Assumpts}

Given the loophole-free experimental violations of Bell inequalities (see Appendix~\ref{sec:Exp}), our ability to violate Bell inequalities must necessarily be due to one of the assumptions used to generate such a Bell inequality being false. We now look in more detail at these assumptions. In this paper, we divide the assumptions into Hidden Variables and Factorisability (which we subdivide into No Superluminal Interaction and Statistical Independence). Since Bell Inequalities are violated by quantum particles, it follows that at least one of these assumptions must be violated. Note there are many other categorisations (e.g. Wiseman and Cavalcanti's \cite{Wiseman2017Causarum}), which subdivide many of these assumptions further (especially the No Superluminal Interaction assumption), or which require us to assume statistical independence before we can use them (e.g., Shimony's Parameter and Outcome Independence \cite{Shimony1993Events})\footnote{{Another recent approach, the quantum reconstruction programme, comes at the problem of Bell inequality violation from a completely different, information-theoretic, or principle (rather than constructive, or physical) perspective, and its proponents claim it shows that such violations can result from an extension of the relativistic principle of ``no preferred reference frames'' to measurements of Planck's constant $h$ \mbox{\cite{Stuckey2020Mermin,Stuckey2022NPRF,Stuckey2024Book}}. However, while interesting, analysis of these results, and how they intersect with the assumptions we analyse here, is beyond the scope of the current paper.}}. However this simple categorisation is sufficient for our purposes, given we are mainly focussing on the statistical independence assumption (which is universal to these categorisations), and how this assumption links to counterfactual definiteness.

\subsection{Hidden variables}\label{subsect:HV}
The first assumption needed for the Bell inequality derivation in Section \ref{subsect:bell} is the existence of hidden, or additional \cite{Maudlin1995,Hance2022Measurement,adlam2023taxonomy}, variables, and the result of a given measurement being a function both of the measurement setting, and some hidden variables. These hidden variables are typically viewed as localised at each particle in some way (i.e. as hidden system properties of each particle). However, this neglects that the hidden variables can also be taken to describe the detectors, and the environment, or may not even be localisable in any meaningful way. Consideration of the ``location'' of hidden variables becomes important when combined with the No Superluminal Signalling assumption mentioned later.
Note, these variables need not completely determine the measurement result (i.e., there can still be some classical randomness to the process of measurement result generation). However, given these variables are hidden, we can just add an additional hidden variable to the model which differentiates between the different random options. This allows us to return to a deterministic relationship between the combination of hidden variables and measurement setting, and measurement results, without affecting the conclusions of the model (see Section 4.2 of \cite{Hance2022Measurement}). 

We can represent this dependence on hidden variables by writing our product result $AB$ as
\begin{equation}\label{eq:HV}
    AB \equiv A(a,b,\lambda)B(a,b,\lambda)
\end{equation}
where $a$ and $b$ are our two measurement setting choices ($a\in\{a_0,a_1\}$, $b\in\{b_0,b_1\}$), and $\lambda$ represents our hidden variables. For greatest generality, we write each measurement result $A$, $B$ as a function of both the measurement choice applied on its side, and the measurement choice applied to the other side. This is restricted further by the Factorisability assumption discussed below.

We can imagine Eq.~\ref{eq:HV} as providing us with a black-box approach to quantum mechanics, where the box has inputs $a$, $b$, and $\lambda$, and outputs $A$ and $B$. We would expect to always be able to pick a $\lambda$ such that we can explain any $A$ and $B$, for all possible $a$s and $b$s. While the $\lambda$s here are often thought of as numbers, we can imagine the $\lambda$s as being any way we can choose to relate $a$s and $b$s to $A$s and $B$s (e.g., matrices, or even higher-dimensional-tensors).

The hidden-variables assumption is often claimed to be rebutted by the violation of Bell inequalities. Those who hold to this interpretation claim that what Bell's Theorem shows is that quantum mechanics provides a complete description of the world, and hidden variable models are unnecessary classical baggage we bring with us. However, given it requires the intersection of the hidden variables assumption with a number of other assumptions to generate a Bell inequality, it is obviously not true that hidden variables are necessarily ruled out by Bell inequality violations. Further, taking the quantum formalism at face value still requires the violation of another key assumption necessary to generate a Bell inequality: factorisability.

\subsection{Factorisability}\label{subsect:Factorisability}
The next assumption, which is necessary to form a Bell inequality, is factorisability. This is the assumption we can write the product of the measurement results in the form
\begin{equation}
    AB \equiv A(a,\lambda_A)B(b,\lambda_B)
\end{equation}
where there are two implicit assumptions. The first is that the hidden variables $\lambda_A$ are only able to be influenced by events in the past or future light cone of the measurement event giving $A$, and the same for $\lambda_B$ and $B$. The second is that the measurement settings $a$ and $b$ can be treated as completely free variables---they are not themselves influenced by each other, or the hidden variables, nor can they influence the hidden variables.

This notion of factorisability therefore can be decomposed more fully into these two assumptions---no superluminal interaction, and statistical independence. 

Given the quantum formalism represents entangled states as the superposition of different tensored states, which cannot be written in such a way that the state of each subsystem is separate (i.e. as the tensor product of some state/superposition of states of particle $A$, and some state/superposition of states of particle $B$), the quantum formalism in some way looks to violate some notion of factorisability. However, this tensor-product-factorisability is formally different to the notion of measurement-result factorisability we give above.

\subsubsection{No superluminal interaction}\label{subsect:NoSuperluminal}
The No Superluminal Interaction assumption is that there can be no way for spacelike-separated events (those outside each others' past or future light-cones) to affect each other. By having the two measurement events spacelike-separated (as is required to avoid the locality loophole), this assumption prevents the measurement setting on one side of the set-up directly affecting the measurement result on the other side of the set-up (as in, affecting the result beyond such correlation as would be allowed by the shared part of the hidden variables). Therefore, it allows us to partially-factorise the product of the measurement results into
\begin{equation}
    AB \equiv A(a(\lambda),\lambda)B(b(\lambda),\lambda)
\end{equation}
However, this does not necessarily mean this assumption requires the measurement settings on one side cannot affect the measurement results from the other, as this correlation between the two could be carried through a correlation between the measurement settings and the hidden variables. (This possibility is only blocked by combining the No Superluminal Interaction assumption with the Statistical Independence assumption below.)

A concern with models which break this assumption is that we haven't observed any other systems where interaction occurs superluminally. Shimony therefore proposed the idea of quantum interactions exhibiting superluminal `passion', rather than action, at a distance \cite{Shimony1993Passion}. This is as, even if factorisability was violated by some superluminal effect for entangled particles, no observer would be able to extract or transmit information superluminally through this mechanism, due to the No-Signalling Theorem \cite{ghirardi1980general,Peres2004QIandRel}. Therefore, the violation of this assumption doesn't necessarily allow observations which would contradict Special Relativity.

That this assumption is the one violated is one of the most common interpretations of Bell's Theorem, to the extent that Bell's Theorem is often claimed to show that nature is nonlocal \cite{NatPhys2022Survey}, and EPR-correlation (or at least the quantum formalism's violation of tensor-product-factorisability) is often referred to in quantum information theory as nonlocality \cite{Popescu1994Boxes,Popescu2014Nonlocality,brunner2014bell}. However, while such an interpretation respects No-Signalling, it still seems to violate our intuitions about Special Relativity, given this prima facie prohibits any signal, not just observable information, travelling superluminally.

\subsubsection{Statistical Independence}\label{subsect:SI}
Statistical independence is the assumption that there are no correlations between our measurement choices and our hidden variables. This can be represented in multiple equivalent ways.

Statistical Independence generally refers to the assumption that hidden variables $\lambda$ are uncorrelated with the choices of measurement settings for a given measurement.

Following the notation we use above, we can write this assumption as
\begin{equation}
    a\neq a(\lambda);\;b\neq b(\lambda);\;\lambda\neq \lambda(a);\;\lambda\neq\lambda(b)
\end{equation}

Alternatively, if we imagine some probability distribution over our hidden variables describing their likelihood, $\rho$, then we can write this assumption as
\begin{equation}\label{SIstandard}
    \rho(\lambda,a,b) = \rho(\lambda),\;\forall a\in\{a_0,a_1\},\,\forall b\in\{b_0,b_1\}
\end{equation}
where $\rho$ is the probability of the hidden variables having a certain value $\lambda$.

It is a common misconception that this assumption is not needed, and so Bell's Theorem rules out all local hidden-variable models \cite{Schlosshauer2013Attitudes,Sujeevan2016Survey,NatPhys2022Survey}. However, the assumption is mathematically necessary to formulate a Bell inequality \cite{chen2021bell,Hance2022ComNatPhys}---otherwise, the measurement result on one side of the experiment could trivially depend on the measurement setting on the other side of the experiment through the hidden variables, which would violate factorisability without requiring superluminal interaction. Therefore it makes sense to consider models which violate this statistical independence assumption.

The amount of statistical independence violation which is necessary to reproduce quantum mechanics is a hotly-debated topic. Hall showed that remarkably little violation of statistical independence is necessary to allow a local hidden variable model to reproduce quantum mechanics \cite{Hall2011Relaxed}, and Kimura et al showed this holds even for a remarkably small space of hidden variables \cite{kimura2023relaxed}. However, Putz and Gisin showed that showing that keeping any arbitrarily small amount of measurement independence (as they define it, $l>0${)}, means some (also potentially arbitrarily small) violation of no-signalling is necessary for a model to reproduce quantum mechanics \cite{Putz2014ArbMI,Putz2016MDL}, and Vieira et al recently showed that a model needs either complete nonlocality or complete measurement dependence $l=0$ in order to reproduce quantum mechanics \cite{vieira2024test}.

We can imagine three cases (ignoring just repeated coincidence) where the statistical independence assumption is not true (i.e., where there is some correlation between the hidden variables and measurement settings). First, this correlation could be due to the hidden variables in some way influencing the choice of measurement settings; secondly, the measurement settings could in some way influence the value of the hidden variables; or thirdly, the choice of measurement settings and value of the hidden variables could have some common cause. Models following the first or third option are commonly referred to as superdeterministic \cite{Hossenfelder2020Rethinking,Hossenfelder2020SuperdeterminismGuide} (although in \cite{Hance2022Supermeasured} and below we discuss another set of models which fall under the third option: supermeasured models). Models following the second option are termed retrocausal: we do not discuss these further in this paper, but recommend \cite{Wharton2019Reformulations} for a thorough review of these models.

While violation of the statistical independence assumption explains the apparently instantaneous, mechanism-free collapse of one particle based on the other (due to measurement choices themselves being correlated with the hidden variables), it has been argued that it presents both epistemic issues (making it impossible to empirically derive physical laws), and physical issues (such as being fine-tuned to the point of conspiracy) \cite{Sen2020Superdet1,Sen2020Superdet2}. While such bold claims have been refuted elsewhere \cite{Hossenfelder2020Rethinking,Hossenfelder2020SuperdeterminismGuide}, such a discussion is not the subject of our paper.

Examples of superdeterministic models include Brans's model \cite{Brans1988Model}, 't Hooft's cellular automaton model \cite{tHooft2016Cellular}, Ciepielewski, Okon and Sudarsky's model \cite{Ciepielewski2020Superdeterministic}, Donadi and Hossenfelder's toy model \cite{Donadi2020SuperdetToy}, and Palmer's Invariant Set Theory \cite{Palmer2020Discretization}. Invariant Set Theory is an example of a model which fits more neatly in a new subcategory of statistical independence-violating model we proposed in \cite{Hance2022Supermeasured}---supermeasured models.

While authors often talk about classical superdeterministic models \cite{Daley2022Adjudicating}, all the models mentioned above use entanglement, superpositions, density matrices and wavefunctions just like standard quantum mechanics \cite{Hance2022DaleyComment}. This illustrates why it is worth investigating local hidden-variable models which avoid Bell's Theorem: not just to return to something which looks like classical mechanics, but to resolve issues which the standard formalism of quantum mechanics cannot. An example of such an issue is the Measurement Problem \cite{Hance2022Measurement}.

\subsubsection{Supermeasured}

An alternative understanding of how statistical independence may be violated, is the application of a supermeasure (a non-trivial measure) to the relation between probability space and state space \cite{Hance2022Supermeasured}. First we must acknowledge the existence of a measure $\mu$ to move from probability space to state space - i.e.

\begin{equation}
    \rho_{\rm Bell} (\lambda|a,b) := \rho(\lambda|a,b)\mu(\lambda|a,b)
\end{equation}
where $\rho_{\rm Bell}$ is the response function on state space for a given hidden variable $\lambda$, used to formulate Bell inequalities.

Therefore, the actual assumption used in Bell's theorem, rather than that given in Eq.~\ref{SIstandard}, is

\begin{equation}
    \rho_{\rm Bell} (\lambda|a,b) = \rho_{\rm Bell} (\lambda)~
\end{equation}

This means, instead of treating the Statistical Independence assumption as a primitive (often referred to as the ``Free Choice'' assumption, as in Wiseman and Cavalcanti \cite{Wiseman2017Causarum}), we should instead view it as the intersection of Eq.~\ref{SIstandard} and 

\begin{equation}
\label{Eq.StatIndMuViolation}
    \mu(\lambda|a,b) = \mu(\lambda),\;\forall a,b,\lambda
\end{equation}

A violation of this implies a non-trivial measure taking us to probability space from state space. {Let us try and put supermeasured models into the terminology of \mbox{\cite{adlam2023taxonomy}}. The first challenge when doing this is whether the measure really counts as an input, given \mbox{\cite{adlam2023taxonomy}} say ``To be part of the inputs, an assumption must differ between at least two scenarios.'' While we could hypothesise worlds where it is (mathematically) different, the nature of the measure in supermeasured models is that it is a physical fact, in the same way as values of physical constants, like $c$ and $\hbar$. Therefore, whether it could be said to ``differ between at least two scenarios'' depends whether you are considering the mathematical possibility or physical possibility of it differing (yes for the first, no for the second). Assuming we accept the measure as an input, by the terminology in \mbox{\cite{adlam2023taxonomy}}, it would be an ``all-in-one'' input, so models of this form would be ``all-at-once'' models, albeit ones typically respecting ``continuity of action'' (otherwise they would be able to violate the ``No superluminal interaction'' condition, so wouldn't also need to violate statistical independence to violate Bell inequalities).}

Palmer \cite{Palmer1995Spin,Palmer2009ISP,Palmer2016IST,Palmer2020Discretization} constructed a model which violates Bell inequalities by proposing non-trivial measures (and also is suggested to be extendable to reproduce General Relativity)\footnote{{One issue with Palmer's Invariant Set Theory in its current form is, that while it can provably violate Bell inequalities (as shown in e.g., \mbox{\cite{Palmer2019Bell}}), it has not been shown what additional constraints are necessary on the Theory for it to violate Bell inequalities in the same way as quantum mechanics (i.e., to the Tsirelson bound, but no further). This presents an interesting direction for future work.}}. However, `supermeasured' models may also imply limits on counterfactual definiteness, as we will discuss below.

\section{Ability to compare cases with different measurement settings}\label{subsect:Comparison}

An additional assumption often mentioned as necessary to form a Bell inequality, is that there is some way to compare scenarios with different measurement settings. This comparison can be done in one of two ways---statistically, or counterfactually. 

\subsection{Statistical Comparison}

We here first look at the statistical method, which is the method typically used experimentally, where results from multiple different Bell tests (different $i$s, to use the terminology above) are combined together.
However, this involves considering probability distributions over possible measurement result pairs, which requires assuming Kolmogorov's axioms \cite{kolmogorov2018foundations}.
When all of these probability distributions are nonzero, as is necessary to represent the EPR-Bohm experiment, these axioms lead to one of two conclusions.
The first option is, the probability distributions could be an incorrect representation of the situation (representing instead the experimenter's incomplete knowledge).
Alternatively, if the distributions provide accurate representations of the situation, these probability axioms place counterfactual definiteness conditions on the four result pairs at least as strong as the Matching Condition. This is as it requires the situation where a certain result pair has a nonzero probability distributions to be meaningful enough to assign nonzero probability to.
Therefore, moving to probabilistic models does not remove the dependency on counterfactual definiteness, or show the Matching Condition to be unnecessary.
However, these axioms are far stronger than the matching condition, and (depending on how one interprets these probabilities) such an approach can itself be said to rely on counterfactual definiteness as a sub-assumption.

However, an issue with this method is it assumes that the hidden variable $\lambda$ will remain the same across different tests (different ``$i$''s). This is a fundamentally unverifiable assumption, given the hidden-ness of the hidden variables $\lambda$ (and this unverifiability provides a rationale for us thinking the Statistical Independence assumption might be violated in these cases). While arguments are often given in analogy to classical experiments, that we should assume that we can hold $\lambda$ fixed, as we can screen off causal influences from the experiment (e.g. isolate labs from the environment, keep apparatus as similar as possible between tests, etc), this often assumes only classical influences, which given we are investigating non-classical correlations, is dubious at best. Bell himself admitted this, saying ``In this matter of causality it is a great inconvenience that the real world is given to us once only... We cannot repeat an experiment changing just one variable; the hands of the clock will have moved, and the moons of Jupiter'' \cite{Bell1964}. This indicates the issue with such a statistical approach, and the reason for instead comparing results counterfactually.

\subsection{Counterfactual Definiteness}\label{CfD}

Informally, counterfactual definiteness, or determinateness, is the idea that there is a definite outcome for what an observable would be, were something done differently (e.g. for Bell tests, a different measurement basis chosen for one of the two spin measurements). While this is commonly confused with determinism - that the universe evolves deterministically, from a set of initial conditions - counterfactual definiteness is not necessarily deterministic. The universe may be probabilistic and counterfactually definite; so long as there are always definite values for all observables in counterfactual scenarios. This is shown by probabilistic Bell inequalities (e.g. the CHSH inequality \cite{CHSH}), which, despite being probabilistic, still gives an inequality which quantum phenomena can violate.

Therefore, being able to violate Bell inequalities could mean, for both the Wigner- and probabilistic forms of these inequalities, that the universe isn't counterfactually definite.

While this has been argued previously (e.g., by Maudlin and Lambare \cite{maudlin2010bell,lambare2021note,Lambare2021CfDef}) that counterfactual definiteness is not an assumption of Bell's Theorem, this is typically as they assumed statistical independence, and that statistical, experimental comparisons led to no issues with statistical independence. As we discuss above, this is a problematic assumption, given assuming statistical independence in real-world experiments requires us to assume, to quote Bell, that ``the moons of Jupiter'' are fixed. Therefore, if we want a way to compare values of observables for different measurement choices which does not have the issues statistical comparison does, we need to make a counterfactual comparison---for which we require some level of counterfactual definiteness, so we can compare different definite values for different measurement choices.

This makes us ask just how reduced counterfactual definiteness needs to be to allow a Bell inequality to be violated while keeping all other assumptions. The answer to this is that Bell inequality violation does not require the removal of all counterfactual definiteness---just the violation of the matching condition.

\subsection{The Matching Condition}

The final assumption required to generate a Bell inequality is the matching condition \cite{redhead1987incompleteness}. This is a strong version of the assumption of counterfactual definiteness---that there is a matter of fact about the results of unperformed measurements. In Section \ref{subsect:bell}, we define $\gamma_i$ for a given trial $i$ by summing the four different pairs of measurement results. However, this requires there to be a matter of fact about what result we would get, had we performed each of the four possible pairs of measurements on the particle pair. For instance, imagine for a given trial that we had measured the two particles with measurement settings $a_0$ and $b_0$, and so got measurement result pair $A_{0,i}B_{0,i}$. To generate a value of $\gamma$ for that trial, we are now required to also imagine what the results of a measurement would be, for that trial, had we instead measured $a_1$ instead of $a_0$ (to get a value of $A_{1,i}B_{0,i}$). Further, we are required to imagine both what the measurement result would be had we measured $b_1$ rather than $b_0$ (to get $A_{0,i}B_{1,i}$), and what would happen if we measured both $a_1$ instead of $a_0$ \textbf{and} $b_1$ instead of $b_0$ (to get $A_{1,i}B_{1,i}$). We can only ever measure one pair of measurement settings on each particle pair, so we have to assume it is meaningful to consider these counterfactual options. Note, this is not the same as assuming that changing measurement settings on one side does not affect the measurement results on the other, which is instead part of the locality assumption above. Nor is it the same as saying we can simultaneously \textbf{measure} $A_{0,i}$ and $A_{1,i}$ (or $B_{0,i}$ and $B_{1,i}$)---just that they have definite values which can be simultaneously considered in those pairs. All the matching condition assumption requires is that the measurement results of unperformed measurements are counterfactually definite for all four pairs of measurement choices:
\begin{equation}
    \forall (a\in\{a_0,a_1\},\;b\in\{b_0,b_1\},\;\lambda\in\Lambda),\;\exists A_iB_i
\end{equation}

From our discussion above, one can see that we either need a statistical comparison assumption, or a counterfactual definiteness assumption (e.g. the Matching Condition). While authors such as Maudlin challenge the necessity of a counterfactual definiteness assumption, as they claim it is ``not in the maths'' \cite{maudlin2010bell}, we see from above that such an assumption appears obviously when we try to construct a Bell inequality. This comparative assumption provides a more intuitive meaning to the formal assumption of statistical independence (which is often viewed as abstract, and so violation of is viewed as conspiratorial and fine-tuned), showing exactly how staitsical independence is necessary for us to construct a Bell inequality.

\section{Counterfactual restriction}\label{sect:CFRestrict}

\subsection{Counterfactual restriction rather than no counterfactual definiteness}\label{subsect:CFRnotCFI}

Given the full specification of a scenario, counterfactual definiteness implies there is a fact of the matter on what would happen in any counterfactual situation. Given the full specification of a counterfactual situation, there is a matter of fact about what you'd get. (Note this is distinct from ``and that definite value is invariant under a given change''.) Think of a world {where we measure observables $A$ and $B$ in a given order}\footnote{{Note, while it may be most natural to think of measuring these in a temporal order, this could be any ordering whatsoever (e.g., spatially, on different particles, going from leftmost to rightmost particle).}}{---specifically, first measure $A$, then measure $B$ (which we can write as $A_1$, $B_2$)}. Counterfactual definiteness means, in this world, {$A_1$} has value $c$. We didn't measure {$A_1$}, but there is a fact of matter on what we would have got had we measured it. Counterfactual indefiniteness is instead where we could have measured {$A_1$}, but there is no fact of matter about what would have obtained. To put these into an example, consider a situation where even though we didn't have lunch, if we had had lunch, we would have had chips rather than rice. Such a scenario is counterfactually definite. If, instead, we can't have said whether would have had chips or rice, then the scenario is counterfactually indefinite.

Instead, a counterfactual restriction is where it is nomologically impossible for {$A_1$} to be measured---even though we can imagine the possibility (it is logically and potentially metaphysically possible), it is not possible under the actual laws of nature \cite{KmentSEPModality}. This occurs when there are things we thought independently vary, but that we cannot independently vary---while we can imagine counterfactual scenarios where one is varied while the other is kept fixed, counterfactual scenarios of this sort are not physically possible. The counterfactual situation doesn't obtain, rather than isn't definite. A practical implication of this is that, once a quantum system is measured one way, it means it couldn't have been measured in certain other ways. For example, in Palmer's Invariant Set Theory, the invariant set's constraint is a restriction on what we can measure. Going back to the lunch example, this is equivalent to saying that, we couldn't have had lunch in first place---it was nomologically impossible.

We can see that one way to violate the matching condition therefore is to link $a$, $b$, and $\lambda$, such that there are triples of $a$, $b$, and $\lambda$ which are counterfactually restricted---which cannot ever be measured, and so $A_iB_i$ does not exist. However, given this requires a link between the measurement settings $a$ and $b$ and the hidden variable $\lambda$, this seems to link the violation of the matching condition to the violation of statistical independence.

\subsection{Counterfactual restriction as subtype of SI violation}\label{subsect:CFRasSIV}

The violation of Eq.~\ref{Eq.StatIndMuViolation} can also be thought of as a form of counterfactual restriction---by having the measure $\mu$ as non-trivial (i.e. having $\mu$ being 0 for certain triples of $a,b$ and $\lambda$), we can say these triples are counterfactually restricted. This, combined with the analysis in Section \ref{subsect:CFRnotCFI}, leads us to ask whether, more generally, counterfactual restriction can be considered instead as a specific sub-type of violation of the statistical independence assumption which factors into Bell's theorem.

To consider this, let us look again at the matching condition. As we say above, it requires the value of all four result products in our CHSH inequality are simultaneously defined for any $\lambda$, and so that any triple of $a$, $b$ and $\lambda$ is nomologically possible. 

If we take a specific state as being defined if it has a nonzero measure on state space, the matching condition is equivalent to saying
\begin{equation}
    \mu(\lambda|a,b)\neq 0,\;\forall (a\in\{a_0,a_1\},\;b\in\{b_0,b_1\},\;\lambda)
\end{equation}

Therefore, a violation of the matching condition is just a case where, for at least one triple of $a$, $b$ and $\lambda$, $\mu$ is equal to 0.

As $\rho_{\text{Bell}}$ is what goes into Bell's theorem, rather than $\rho$, this still allows us to maintain physical statistical independence (or to, say, for any given triple of $a$, $b$ and $\lambda$, calculate the expectation value for each result product)---it just means we cannot combine the four different expectation values together to form a Bell inequality.

Despite being a violation of what formally enters Bell's theorem as the Statistical Independence assumption, this isn't what most people think of when they think of statistical independence being violated---these cases can still obey
\begin{equation}
    \rho(\lambda|a,b)=\rho(\lambda),\;\forall (a\in\{a_0,a_1\},\;b\in\{b_0,b_1\},\;\lambda)
\end{equation}
and can even still obey Bell-Statistical Independence for all allowed counterfactual situations---that is
\begin{equation}
\begin{split}
    &\rho_{\text{Bell}}(\lambda|a,b)=\rho_{\text{Bell}}(\lambda),\\
    &\forall (a\in\{a_0,a_1\},\;b\in\{b_0,b_1\},\;\lambda)\;\text{where}\;\mu\neq0
\end{split}
\end{equation}

The fact that violations of the matching condition can be represented as violations of Bell-Statistical Independence (but not physical statistical independence), or to use other terminology, as a form of supermeasured theory, makes us ask whether we should reassess how we typically consider statistical independence-violating models.

Let us look at the archetypal supermeasured theory---Palmer's {I}nvariant {S}et {T}heory.

In Palmer's Invariant Set Theory, there is a clear (but uncomputable) split between those states which are on the invariant set, and so allowed by the theory, and those which are not on the set, and so are prohibited by the theory. This split occurs in such a way that, if a pair of states exist such that each state has a definite value for one of a pair of conjugate variables (e.g. one state has a definite value for spin in the $x$-direction, and the other a definite value for spin in the $z$-direction), the two states will never both be on the invariant set, and so at least one of the two states will never be a (nomologically) possible state.
This restriction on states with definite values of conjugate variables both being on the invariant set looks in some sense like a counterfactual restriction on the pair of states for a given value of $\lambda$---if one of the pair is counterfactually definite, the other by definition is counterfactually restricted.

Another set of models which look to be cases of a counterfactual restriction are those described as having an epistemic restriction, such as Spekkens' Toy Model \cite{Spekkens2007Toy}, and those models derived from it \cite{vanEnk2007,Paterek2010Limited,Coecke2011Toy,Bartlett2012Liouville,Spekkens2016Quasi}. While these models often don't fully reproduce quantum mechanics, they reproduce certain properties of quantum mechanics despite being nominally classical, through the application of an epistemic restriction. Given the epistemic restriction serves to change the behaviour of the system in ways far stronger than one would normally attribute to limiting knowledge about a system, the ``knowledge balance principle'' underlying these models looks more like a counterfactual restriction than a real restriction on knowledge. Interpreting the principle this way, these models also look to show that quantum features can be regained by a counterfactually-restricted classical model.

Such a counterfactual restriction also looks to be an explanation for the observable effect of measurement-induced back-action, where performing a measurement on a quantum system has a measurable effect, even if the measurement was ``interaction-free'' (there was no direct path of interaction between the system, typically a quantum particle, and the measuring apparatus, typically a blocker/absorber). This is discussed further in \cite{Hance2024CFBAIG}, and the implications of this will be explored in future works.

\subsection{Links to Contextuality}\label{subsect:Contextuality}

The assumption going into Bell's Theorem that we are free to choose all variables we view as ``free'' classically (i.e. statistical independence, or that there is no counterfactual restriction) ties into the burgeoning research area of contextuality theory \cite{Budroni2022ContextualityReview}. This investigates into the idea that the key difference between classical and quantum phenomena is that classical phenomena are always the same regardless of measurement context, whereas quantum phenomena are context-dependent. While the theory of contextuality has been tied into statistical independence violation previously \cite{Dzhafarov2015,Dzhafarov2016CbDReview}, this is (surprisingly) a fringe position. This is despite statistical independence being the assumption that the choice of measurement (i.e. measurement context) plays no effect on a quantum system, and so models being contextual (where quantum systems are dependent on measurement contexts) and models violating statistical independence (where quantum systems are dependent on measurement settings used) being obviously equivalent. This cognitive dissonance between contextuality being accepted but statistical independence violation being rejected may be due to the typical scenarios these two model behaviours are demonstrated in: a system depending on the measurement context (the totality of measurement settings for the scenario) is possibly more palatable in Kochen-Specker-style scenarios (where spatial separation between elements of the measurement context isn't emphasised) than in Bell scenarios (where spatial separation between elements of the measurement context \emph{is} emphasised). The former seems more obviously some form of restriction of possibilities given the context, whereas the latter seems to push for some realist physical mechanism whereby the separate parts of the measurement context (in the EPR-Bohm scenario, the measurement choices at Alice and Bob) coordinate with one another. This links to typical claims that statistical independence violating models are in some sense fine-tuned, or conspiratorial, even when those terms are not well-defined. However, considering statistical independence violation instead as a restriction on counterfactuals, rather than requiring a conspiratorial physical mechanism, both helps resolve this worry and reiterates the link between SI-violation and contextuality (which makes sense, given the EPR-Bohm scenario, and any other Bell Inequality-violating scenario, is also inherently contextual).

\section{Conclusion}\label{sect:Conc}

We introduced Bell's Theorem, and looked at both the assumptions necessary to formulate a Bell inequality, and the experimental loopholes which can allow models which still meet all these assumptions to violate such an inequality.

By looking at the assumptions underpinning Bell inequalities, we showed that one way of interpreting the theorem is that some form of counterfactual restriction allows a model where, despite all other assumptions necessary to create a Bell inequality being allowed, quantum correlations between measurement results can occur. We linked this counterfactual restriction to statistical independence-violating models (especially those coming under the recently-introduced sub-categorisation of supermeasured models), and then linked both counterfactual restriction and statistical independence-violation to contextuality. This serves not only to undermine claims like Maudlin's that counterfactual definiteness/the matching condition are not necessary to generate Bell inequalities, as they don't appear in the maths (they do, albeit through the statistical independence assumption), but also provides a way of interpreting statistical independence-violating models at odds with their typical conception as either superdeterministic or retrocausal. We hope this helps motivate investigations of links between contextuality, statistical independence violation, and counterfactuality in quantum foundations.

\textit{Acknowledgements -} I thank Sophie M.N. Inman, James Ladyman, John Rarity, Sabine Hossenfelder, Tim N Palmer, and Holger F Hofmann for useful comments. JRH was supported by Hiroshima University's Phoenix Postdoctoral Research Fellowship, the University of York's EPSRC DTP grant EP/R513386/1, and the UK Quantum Communications Hub (EPSRC grants EP/M013472/1 and EP/T001011/1). 

\bibliography{ref.bib}

\appendix

\section{Experimental Tests and Loopholes}\label{sec:Exp}

In this Appendix, we review the history of experimental attempts to violate Bell inequalities, the loopholes present in the arguments supporting early attempts, and how experiments were finally performed which closed these loopholes. The first experimental attempt to violate a Bell inequality was performed by Freedman and Clauser in 1972 \cite{Freedman1972}. This used polarisation-entangled photon pairs, generated through the atomic cascade of Calcium, to try and violate the CHSH inequality that Clauser et al had proposed two years earlier. However, this experiment left a number of loopholes---ways a local hidden-variable theory could still explain the correlations observed, given practical details of the experiment. In this Subsection, we discuss these loopholes, and how later Bell tests managed to avoid them, such that loophole-free violations of Bell inequalities have now been performed. We follow Larsson et al's categorisation of these loopholes into the Locality loophole, the Memory loophole, the Detection loophole, and the Coincidence loophole \cite{larsson2014loopholes}.

\subsection{The Locality Loophole} If local (i.e., luminal or subluminal) communication between the two measurement sites is possible in the time between a measurement choice being decided at one measurement site, and the measurement being performed at the other measurement site, then it is perfectly consistent that we could observe correlations which violate our Bell inequality even if the Universe obeys a local hidden-variable theory. This was the Locality loophole found in Freedman and Clauser's experiment---we could imagine some mechanism by which the measurement settings at the first site (the polariser direction being set) locally affect the state of the particle at the second site, in such a way as to generate these correlations. Note this loophole is also behind supposed macroscopic violations of Bell inequalities, such as those presented by Aerts et al \cite{Aerts2000}, which serve to show very little except an inequality based on an assumption of separation between two subsystems can easily be violated if the two subsystems are not in fact separate. This loophole was first closed in Aspect et al's tests, proposed in 1976 \cite{Aspect1976Proposal}, and performed in the early 1980s \cite{Aspect1981BellViol,Aspect1982}. These experiments also used polarisation-entangled single photons, but used time-varying analysers to perform measurement. These analysers are effectively variable polarisers, which jump between two polarisation orientations in a time far shorter than the time it takes each photon to travel from the source to its respective polariser. 

\subsection{The Memory Loophole} While Aspect et al's experiment closed the locality loophole, it raised a new loophole---the memory loophole. This loophole comes from the analysers in Aspect et al's experiment shifting between the two polarisations deterministically, at a fixed frequency---the state of the polariser at some given time in the future was predictable. While unlikely, one could imagine a way the system could extrapolate what the state would be at a given time to allow the generation of Bell inequality-violating coincidences. Aspect himself noted this loophole in his proposal \cite{Aspect1976Proposal}, saying the set-up required the supplemental assumption that the polarisers have no ``memory'' as to what state they had been in previously, so the system couldn't extrapolate from any regularity in the settings what their state would be in the future. This assumption however is not needed (and the memory loophole closed) if the measurement settings for each experimental run are chosen in some unpredictable way, so there is no way to infer what the measurement setting will be for a given run. The first Bell tests to close this loophole were those of Weihs et al \cite{weihs1998violation}. Here, random number generators at each measurement site determined which of the two possible measurement settings to use for a given run, after the photons were emitted from the source. (This also avoids the locality loophole.)

\subsection{The Detection Loophole} Next, we look at the detection, or efficiency loophole, whereby the possibility of losing or not detecting particles during the experiment can affect the correlations observed. This can occur in two ways. Firstly, when using polarisers (like Freedman and Clauser), one of the measurement results (-1) is recorded when there is no detection, as one assumes the photon, having had an orthogonal polarisation to the polariser's measurement setting, was absorbed by the polariser. However, there are many other reasons the photon might be lost during the experiment. For instance, the detector might not have perfect efficiency, the lenses and transmissive optics between the source and the polariser might not be perfectly transmissive, or polarisers might not be perfectly transmissive even at their transmission polarisation. Therefore, -1 would be recorded far more often in practice than we would expect with a theoretically perfect experiment. This could overshadow any correlation from entanglement, and lead to the data not violating the inequality. To avoid this (and to compensate for not receiving a click when observing a -1 detection), Freedman and Clauser adapted the CHSH inequality to give an inequality in terms of rates of coincidence detection as a function of angle between the two polarisers, divided by rate of coincidences when both polarisers were removed. This however relies on an assumption of fair sampling---that the detection efficiency is the same for different polarisations. This could be violated if, for instance, the detectors had a bias towards detecting certain polarisations more efficiently than others, or the lenses used subtly act as polarisers.

The detection loophole also appears in Aspect et al's experiments. These use polarising beam splitters of variable basis rather than variable polarisers, so there is a detection associated with both a +1 and a -1 click. However, they still neglects runs (photon pairs) where one or both photons are absorbed. Therefore, forming a Bell inequality for these experiments still requires some assumption of fair sampling---that observed runs are a fair sample of all runs. This loophole is only closed by tests where this loss can be bounded and accounted for. Initially, this was only possible in tests using solid-state platforms rather than optics, such as Rowe et al's \cite{Rowe2001BellTest}, which then suffered from the locality loophole. However recent experiments have been performed which close both detection and locality loopholes simultaneously---e.g. Hensen et al's \cite{Hensen2015BellTest}, which use a combination of optics and nitrogen-vacancy centre qubits, and Giustina et al's \cite{GiustinaLoophole2015} and Shalm et al's, which both use entangled photon pairs.

\subsection{The Coincidence Loophole} A related loophole is the coincidence loophole: given we rely on simultaneous detection to tell us which particles where initially generated in the same pair, we could imagine a local hidden variable model whereby state/setting-dependent delays cause correlations which violate the Bell inequality \cite{larsson2004bell}. This loophole is most problematic in photon-pair experiments, such as Freedman and Clauser's, and Aspect et al's, but can again be mitigated in solid-state experiments. In these, it is easier to assign measurement results to a given run, and so they don't require coincidence detection. Larsson showed that later photonic Bell tests (e.g., Giustina et al's and Shalm et al's) also avoid this loophole \cite{Larsson2014Coincidence}.

Bell tests have been performed which close all these loopholes. These experiments show one of the assumptions which go into generating a theoretical Bell inequality must be false, rather than there potentially being some convoluted experimental reason why we observe such a violation.

\end{document}